\documentclass[pre,twocolumn,letterpaper,superscriptaddress,showpacs,floatfix]{revtex4}
\usepackage{amssymb,amsmath,setspace,bm}
\usepackage[dvips]{graphicx}
\usepackage[dvips]{color}

\begin{document}

\title{Critical adsorption and Casimir
torque in wedges and at ridges}

\date{\today}

\author{G. Pal\'agyi}
\email{palagyig@almos.vein.hu}
\altaffiliation{Permanent address: Department of Physics, University of Veszpr\'em, 
H-8201 Veszpr\'em, POBox 158, Hungary} 
\author{S. Dietrich}
\affiliation{Max-Planck-Institut f\"ur Metallforschung, Heisenbergstr. 3,
D-70569 Stuttgart, Germany, and}
\affiliation{
Institut f\"ur Theoretische und Angewandte Physik,
Universit\"at Stuttgart, Pfaffenwaldring 57, D-70569 Stuttgart, Germany}

\begin{abstract}

Geometrical structures of confining surfaces profoundly influence the 
adsorption of fluids upon approaching a critical
point $T_c$ in their bulk phase diagram, i.e., for
$t=(T-T_c)/T_c \rightarrow \pm 0$. Guided by general scaling considerations,
we calculate, within
mean-field theory, the temperature
dependence of the order parameter profile in a wedge with opening angle
$\gamma<\pi$ and close to a ridge ($\gamma>\pi$) for $T\gtrless T_c$ and in
the presence of surface fields.
For
a suitably defined reduced excess adsorption
$\overline{\Gamma}_\pm(\gamma,t\rightarrow \pm 0)
\sim\overline{\Gamma}_\pm(\gamma)|t|^{\beta-2\nu}$
we compute the universal amplitudes $\overline{\Gamma}_\pm(\gamma)$, 
which diverge as 
$\overline{\Gamma}_{\pm}(\gamma\rightarrow 0) \sim 1/\gamma$ for small
opening angles, vary linearly close to $\gamma=\pi$ for $\gamma<\pi$, and
increase exponentially for 
$\gamma\rightarrow 2\pi$. There is evidence that, within
mean-field theory, the
ratio $\overline{\Gamma}_+(\gamma)/\overline{\Gamma}_-(\gamma)$ 
is independent of $\gamma$. We also discuss the critical Casimir torque 
acting on the sides of the wedge as a function of the 
opening angle and temperature.

\end{abstract}

\pacs{64.60.Fr, 68.35.Rh, 68.43.Fg, 61.20.-p}

\maketitle

\section{Introduction}

Boundaries induce deviations of the local structural properties of condensed
matter from their corresponding bulk values. Typically the width of such
boundary layers is proportional to the bulk correlation length $\xi$. Near a
continuous bulk phase transition at a temperature $T=T_c$, the correlation
length diverges according to a power law $\xi_{\pm}(t=(T-T_c)/T_c\rightarrow
\pm 0)=\xi_0^{\pm}|t|^{-\nu}$ with a universal bulk exponent $\nu$ and
nonuniversal amplitudes $\xi_0^{\pm}$ whose ratio $\xi_0^+/\xi_0^-$ is
universal, too. For the three-dimensional Ising universality class
$\nu\simeq 0.63$ and $\xi_0^+/\xi_0^-\simeq 2$ \cite{zinn}, whereas within mean-field
theory, which is valid for spatial dimensions $d\ge 4$, $\nu=1/2$ and
$\xi_0^+/\xi_0^-=\sqrt{2}$. The nonuniversal amplitudes $\xi_0^{\pm}$ are
typically in the order of the range of the interaction potential of the
ordering degrees of freedom, i.e., a few \AA. The correlation length is
defined as the scale of the exponentially decaying two-point correlation
function $\sim e^{-r/\xi}$.

The local critical properties near planar confining surfaces have been
studied theoretically and experimentally in detail. It has turned out, that
each bulk universality class splits up into three possible surface
universality classes denoted as ordinary, special, and normal transitions.
Each of them gives rise to distinct surface critical phenomena. The type of
boundary conditions determines which surface universality class a given
system belongs to. The ordinary transition requires the absence of surface
fields, the special transition is characterized by the absence of a
surface field, too, but also by suitably tuned, enhanced surface couplings
between the ordering degrees of freedom, and the normal transition
represents systems whose surfaces are exposed to surface fields. As
indicated by the name, the latter one is the generic case. For example the
normal transition
applies to one-component fluids near their liquid-vapor critical point, or
to binary liquid mixtures near their demixing critical point. In both cases
the substrate potential of the confining container walls and the missing
interactions due to the fact that the fluid particles cannot penetrate the
substrate give rise to the effective surface fields acting on the
corresponding order parameter (density or concentration difference). As a
consequence, in this case even above $T_c$ there is a nonvanishing order
parameter profile, which decays to zero for $T\ge T_c$ upon approaching the
bulk, i.e., for increasing normal distance $z\rightarrow \infty$ from the
surface at $z=0$. For $T<T_c$ this profile attains the nonzero value of the
bulk order parameter for $z\rightarrow \infty$. Following early calculations
on magnetic systems by K. Binder and P. C. Hohenberg \cite{binhoh}, 
this so-called critical
adsorption was first examined in detail by Fisher and de Gennes
\cite{fisherdegennes} and has since been analyzed for many systems both
theoretically \cite{binder,diehl,diehlrevmod}
and experimentally \cite{smithlaw},
and a fair agreement between theory and experiment has been found
\cite{floedie,sdl}.

Surface order above the bulk critical temperature can also be due to
spontaneous symmetry breaking caused by enhanced surface couplings between
the ordering degrees of freedom, e.g. spins
in magnetic systems. The transition to this state from the disordered state
is usually denoted as the extraordinary transition. Bray and Moore
\cite{braymoore} predicted an equivalence between the normal and the
extraordinary transitions that was later proved by Burkhardt and Diehl
\cite{burkdiehl}. In contrast to the ordinary transition in magnetic
systems, the corresponding extraordinary transition has been
investigated to a lesser extent. Extending and improving earlier work
\cite{leiblerstb,ohno,ciachdiehl,fisherupton}, 
Diehl and Smock \cite{diehlsmock} have
carried out a field-theoretic renormalization-group calculation in
$4-\epsilon$ dimensions for the extraordinary transition in semi-infinite
systems belonging to the Ising universality class \cite{mccoy} 
computing the order
parameter profile to one-loop order. The results are in fair agreement with
those obtained by Monte Carlo simulations \cite{sdl}.

These studies are devoted to the case of planar surfaces.  Only very
carefully treated solid surfaces are atomically flat; generically, however,
they exhibit corrugations. Besides these random deviations from the flat
topography, there is nowadays an abundance of experimental techniques
\cite{letrehozkis} that
allow one to endow solid surfaces with precise lateral geometrical
structures, ranging from the $\mu m$ scale down to tens of $nm$. Inter alia,
such solid surfaces are used within the context of microfluidic devices
\cite{microflu} in order to guide fluids along such structures. The fluids
perfectly fill this laterally structured environment and thus fully exhibit
the strong structural changes associated with that. In this context it is of
interest what kind of structures appear in the fluid if it is brought, say,
close to a demixing transition and is exposed to geometrically structured
substrates. If the characteristic sizes of the lateral structures are
comparable with the correlation length, which in fluids reaches typically up
to a few thousand {\AA} close to $T_c$, one can expect a strong influence on
the aforementioned critical adsorption phenomena.

Theoretically this raises the issue of how the local critical properties
depend on the {\em shape} of the boundaries. A typical manmade structure is
a series of grooves with
various shapes of the cross section, e.g. wedgelike. The first step in their
investigation is the study of a single wedge, which already shows new
features and gives new insight into the influence of geometry on critical
behavior. The ordinary transition of the isotropic N-vector model at
an edge has been
investigated by Cardy \cite{cardywedge} within the framework of mean-field
theory, the renormalization group, and $\epsilon$-expansion. Subsequently
the two-dimensional wedge (corner geometry) was studied by exact
calculations, mainly for Ising models \cite{abraham}, 
and by conformal mapping at the bulk critical
temperature \cite{conformwedge}. New edge and corner exponents were found that
depend on the opening angle $\gamma$ of the wedge. 
More recently similar findings of Monte Carlo simulations 
of the ordinary transition of the three-dimensional Ising model 
\cite{pleimling1} have been reported, in accordance with high temperature
series expansions \cite{hightemp}. The angle dependence of the critical
wedge exponents can be attributed to the fact that the wedge geometry lacks
a length scale and thus is invariant under rescaling. The opening angle is
therefore a marginal variable in a renormalization transformation, and {\em
may} enter into the expressions for the exponents. The same will happen for
all other scale-invariant shapes of the boundaries
\cite{igloi}. However, for a given opening angle, the values of the critical
exponents are expected to be universal and independent of microscopic
details. According to recent Monte Carlo simulations of three-dimensional
Ising models with edges and corners \cite{pleimling2}, angle
    dependent critical exponents are observed at  the ordinary transition,
whereas the surface transition seems to be nonuniversal. The
critical exponents in this latter case appear to depend on the strengths of the
local couplings, in analogy with exact results obtained for the 
two-dimensional Ising model with defect lines
\cite{bariev}. Critical adsorption has also been studied
in general dimensions $d$ on curved surfaces \cite{hankedietrich}.

Since critical adsorption changes the composition of a binary liquid
mixture close to its surface, mechanical properties, such as the local
viscosity and the mutual diffusion coefficient will also vary in space. So
we expect various phenomena, such as flow
properties in porous media, the spreading properties of droplets, surface
chemical reactions, the permeability of membranes etc. \cite{adamson},
  to be influenced significantly by critical adsorption. 

Critical adsorption in a wedge has been studied by Hanke et. al.
\cite{hankestb}. Within mean-field theory the order parameter profile was
calculated exactly {\em at} the critical point. Through an interpolation
scheme between exact results in $d=2$ and the mean-field results
corresponding to $d=4$, angle dependent critical exponents of the order
parameter and those governing the decay of two-point correlation functions
were obtained for $d=3$ as well. The present study extends these
investigations into various directions.

First, we analyze the temperature dependence off criticality for critical
adsorption in a wedge. General scaling properties for the order parameter
profile and the excess adsorption are discussed and the corresponding
scaling functions are calculated within mean-field theory. This analysis is
carried out above and below $T_c$, including a systematic study of the
dependence on the opening angle $\gamma$ of the wedge. This covers also the
case $\gamma>\pi$ describing critical adsorption near a ridge.

Second, as a new feature, we study the Casimir torque acting on the sides of
the wedge or ridge. The confinement modifies the fluctuation spectrum of the
critical fluctuations and the  order parameter profiles. This leads to a
dependence of the free energy of the critical medium on the shape of and the
distance between the confining boundaries, which results in an effective
force acting on them. Thus the physical origin of this force, originally
predicted by Fisher
and de Gennes \cite{fisherdegennes} for two parallel plates immersed into a
binary liquid mixture near its continuous demixing transition, is analogous
to the Casimir force acting on conducting plates in vacuum due to the
confinement of
quantum mechanical vacuum fluctuations of the electromagnetic field
\cite{casimireredeti}. The Casimir force is governed by universal scaling
functions \cite{krech} and is superimposed on the noncritical background
forces, which in the case of fluids are given by dispersion forces. Recent
experiments \cite{chan} have confirmed corresponding theoretical predictions
for the plate geometry \cite{krechdietrich}. For curved surfaces the
critical Casimir force plays an important role in the flocculation of   
colloidal particles suspended in a solvent undergoing a continuous phase
transition \cite{colloid}. In the present context the free energy of the
critical medium depends on the opening angle $\gamma$; its derivative with
respect to $\gamma$ amounts to the critical Casimir torque acting on the
sides of the wedge or ridge. If the substrate forming the wedge or ridge is
composed of soft materials like, e.g., membranes, this critical Casimir torque
is expected to give rise to elastic deformations. It might also be experimentally
accessible by suitable force microscopy with moveable sidewalls of a wedgelike 
structure.

The paper is organized as follows. In Sec. II we discuss the general scaling 
properties of the order parameter profiles and the excess
adsorption. The scaling functions are determined within mean-field
theory and analyzed in detail in Sec. III. Section IV focuses on the free
energy of the confined fluid and the critical Casimir torque resulting
from its angle dependence. Section V summarizes our findings. In the
Appendix we discuss how the excess adsorption in a wedge or at a
ridge decomposes into surface and line contributions
with two possible experimental realizations.


\section{General scaling properties of order parameter profiles and excess
adsorption}

Since fluids can fill a container of arbitrary shape, in the present context
of critical systems exposed to substrates shaped as wedges we consider {\em
fluids} 
 close to their bulk critical point $T_c$. This can be
either a liquid-vapor critical point or a demixing critical point in the
case of binary liquid mixtures. The interaction of the container walls with
the fluid particles results in a spatial variation of the number densities
close to the boundaries.  The deviation of the density of the fluid, or of
the concentration of one of its two components in the case of binary liquid
mixtures, from the corresponding bulk value at $T_c$ is chosen as the local
order parameter describing the phase transition.

The order parameter profile $m^{\infty/2}_\pm (\zeta,t)$ near a {\em planar} interface and 
close to the critical temperature $T_c$ takes the following
scaling form \cite{floedie,sdl,ciachdiehl,diehlsmock,diehlber}:
\begin{eqnarray}
\begin{array}{ll}
m^{\infty/2}_\pm (\zeta,t)=a|t|^\beta P^{\infty/2}_\pm(\zeta/\xi_\pm), & t=(T-T_c)/T_c,
\end{array}
\label{skalaforma}
\end{eqnarray}
for distances $\zeta\gtrsim\sigma$ perpendicular to the interface and
sufficiently large compared to a typical microscopic length $\sigma$. 
$\xi_\pm (t\rightarrow
0)=\xi_0^\pm |t|^{-\nu}$ is the bulk correlation length above ($+$) or below
($-$)
$T_c$, $\beta$ and $\nu$ are the standard bulk critical exponents, and 
$t$ is the reduced temperature.
The scaling functions
$P^{\infty/2}_\pm(\zeta_\pm=\zeta/\xi_\pm)$ 
are universal once the nonuniversal bulk amplitudes $a$ and
$\xi_0^\pm$ are fixed, where $a$ is the amplitude of the bulk order
parameter $m^{\infty/2}_-(\zeta=\infty,t\rightarrow 0^-)=a|t|^\beta=m_b(t)$.
 With the prefactors
$\xi_0^\pm$ fixed as those of the {\em true} correlation length defined by
the exponential decay of the bulk
two-point correlation function in real space,
one finds $P^{\infty/2}_-(\infty)=1$, $P^{\infty/2}_+(\infty)=0$,
$P^{\infty/2}_-(\zeta_-\rightarrow
\infty)-1\sim e^{-\zeta_-}$, $P^{\infty/2}_+(\zeta_+\rightarrow \infty)\sim 
e^{-\zeta_+}$ and
$P^{\infty/2}_\pm(\zeta_\pm \rightarrow 0)=c_\pm \zeta_\pm^{-\beta/\nu}$ 
\cite{diehlsmock}, so
that
\begin{equation}
m^{\infty/2}(\zeta,t=0)=ac_\pm(\zeta/\xi_0^\pm)^{-\beta/\nu}.
\label{limitbehav}
\end{equation}
Any other choice for the definition of the correlation length leads to a
redefinition of the scaling functions $P^{\infty/2}_\pm$ such that all observable
quantities remain unchanged. This underscores that the scaling functions are
universal, but that their form depends on the definition of the correlation
length. The amplitudes of the scaling functions are fixed by the requirement
$P^{\infty/2}_-(\infty)=1$. Accordingly the numbers $c_\pm$ are universal surface
amplitudes which are definition-dependent \cite{floedie}.

Close to $T_c$ the total enrichment at
the interface of, say, the A particles as compared to the B particles of a
binary liquid mixture is
given by the excess
adsorption, which is an experimentally accessible integral quantity. For a
planar surface, one has
\begin{equation}
\Gamma^{\infty/2}_\pm (t)=\int_0^\infty [m^{\infty/2}_\pm(\zeta,t)-
m^{\infty/2}_\pm(\zeta=\infty,t)] d\zeta
\label{adszdef}
\end{equation}
The scaling behavior of $\Gamma^{\infty/2}_\pm (t)$ has been discussed in
Ref. \cite{floedie} in detail. The $\zeta$ integration can be split into the
intervals $\zeta>\sigma$ and $0\le\zeta\le\sigma$, and  for large $\zeta$ 
the order parameter profile 
$m^{\infty/2}_\pm(\zeta,t)$ can be replaced by Eq. (\ref{skalaforma}), which gives
\begin{multline}
\Gamma^{\infty/2}_\pm (t)=\int_0^\sigma [m^{\infty/2}_\pm(\zeta,t)-
m^{\infty/2}_\pm(\zeta=\infty,t)] d\zeta
\\
+a\xi_0^\pm|t|^{\beta-\nu}\int_{\sigma/\xi_\pm}^\infty
[P^{\infty/2}_\pm(\zeta_\pm)-P^{\infty/2}_\pm(\zeta_\pm=\infty)] d\zeta_\pm.
\label{sikskala}
\end{multline}
Upon approaching $T_c$ the first integral remains finite and yields a
nonuniversal constant, which is subdominant to the diverging second term.
The second integral leads to the well known power-law singularity of 
$\Gamma^{\infty/2}_\pm(t\rightarrow 0)$ for $d<4$:
\begin{equation}
\Gamma^{\infty/2}_\pm (t\rightarrow 0)=a\xi_0^\pm g_\pm
\frac{|t|^{\beta-\nu}}{\nu-\beta}, \qquad d<4,
\label{renormalando}
\end{equation}
where the numbers $g_\pm$ are universal with their values depending on the
definitions of the bulk order parameter and the correlation length. In $d=4$
one finds upon inserting the mean-field expression for $P^{\infty/2}_\pm$ (see
Ref. \cite{diehlsmock}) into Eq. (\ref{sikskala}) 
that $\Gamma^{\infty/2}_\pm (t)$ diverges logarithmically. This result can be
reconciled with Eq. (\ref{renormalando}) by noting that 
$\Gamma^{\infty/2}_\pm$
actually needs additive renormalization leading to \cite{floedie}
\begin{equation}
\Gamma^{\infty/2}_\pm (t\rightarrow 0)=a\xi_0^\pm g_\pm
\frac{|t|^{\beta-\nu}-1}{\nu-\beta}, \qquad d\le 4.
\label{renormaltskala}
\end{equation}

\begin{figure}[!h]
\includegraphics*[width=0.8\linewidth]{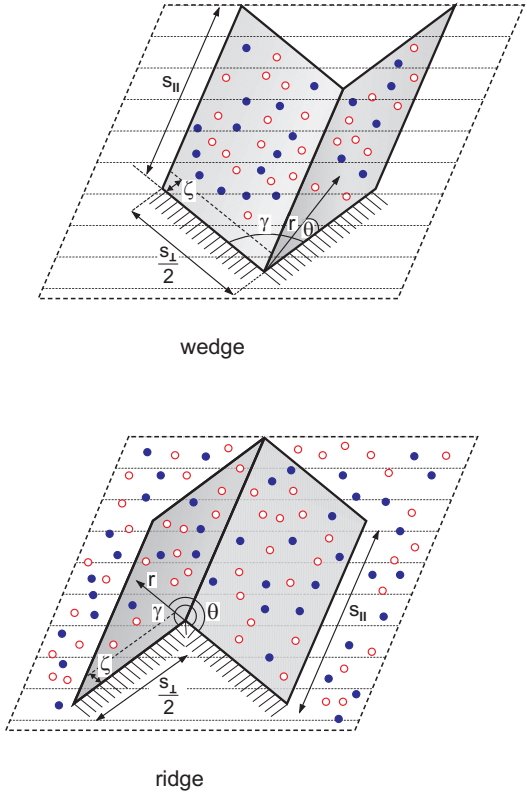}
\caption{A wedge and a ridge with opening angle $\gamma$ exposed to
 a binary liquid mixture. 
The system is translationally invariant in the subspace
parallel to the edge. Within a plane orthogonal to the edge the polar
coordinates are $r$ and $\theta$. The linear extensions of the confining
surfaces are $s_\parallel$ and $s_\perp$.}
\label{abra:ek}
\end{figure}

In the {\em wedge or ridge geometry} (Fig. \ref{abra:ek}) the order
parameter profile depends on the radial distance $r$ from the apex, on the
polar angle $\theta$, and on the opening angle $\gamma$, so that the variation 
of the profile is two-dimensional with corresponding generalized scaling
functions: 
\begin{equation}
m_\pm(\bm{r},t;\gamma)=a|t|^\beta P_\pm(r/\xi_\pm,\theta;\gamma).
\label{skalaekprof}
\end{equation}
As before, the scaling functions are universal once the nonuniversal bulk 
amplitudes $a$ and
$\xi_0^\pm$ are fixed, where $a$ is the amplitude of the bulk order
parameter and the amplitudes 
$\xi_0^\pm$ are prefactors of the {\em true} correlation length 
as in the case of the infinite planar wall.
One finds \cite{hankestb} with $r_\pm=r/\xi_\pm$ 
that $P_-(r_-=\infty,\theta)=1$, 
$P_+(r_+=\infty,\theta)=0$,
$P_-(r_-\rightarrow\infty,\theta)-1\sim e^{-\sin(\theta)r_-}$
($\theta<\gamma/2$),
 $P_+(r_+\rightarrow \infty,\theta)\sim e^{-\sin(\theta)r_+}$
($\theta<\gamma/2$),
$P_\pm(r_\pm \rightarrow 0,\theta)=\tilde{c}_\pm(\theta,\gamma) 
r_\pm^{-\beta/\nu}$,
and 
\begin{equation}
\begin{split}
P_\pm(r_\pm,\theta\rightarrow 0)&=c_\pm \zeta_\pm^{-\beta/\nu}\\
&=c_\pm (r_\pm\sin\theta)^{-\beta/\nu},
\label{limitbehavek}
\end{split}
\end{equation}
where $\zeta_\pm=r_\pm\sin\theta$ (see Fig. \ref{abra:ek}).
 The amplitudes of the scaling functions are again fixed by the requirement
$P_-(r=\infty,\theta)=1$. The numbers $c_\pm$ are the universal surface
amplitudes of the scaling function of the infinite planar wall [Eq.
(\ref{limitbehav})], 
and $\tilde{c}_\pm(\theta,\gamma)$ are universal functions. Both $c_\pm$ and
$\tilde{c}_\pm$ depend on the
definition of the correlation length. 
The scaling functions also reflect the
symmetry of present the geometry: 
\begin{equation}
P_\pm(r_\pm,\theta;\gamma)=P_\pm(r_\pm,\gamma-\theta;\gamma).
\label{symmetry}
\end{equation}

We define the {\em excess} adsorption for this geometry confined by surfaces
 of linear extensions $s_\perp$ in the plane
perpendicular to the edge and $s_\parallel$ in the translationally 
invariant directions (see Fig. \ref{abra:ek}) as
\begin{equation}
\tilde{\Gamma}_\pm(s_\perp,s_\parallel,t;\gamma)=\int_V
d^dr[m_\pm(\bm{r},t;\gamma)-m_b(t)],
\label{gammahullam}
\end{equation}
where the integral is taken over a macroscopic volume $V$ 
occupied by the liquid. 
According to the Appendix this excess adsorption decomposes into a 
surface contribution that scales with the
actual surface area of the confining walls 
($s_\perp s_\parallel^{(d-2)}$) and a line 
contribution that scales with the extension in the invariant directions
($s_\parallel^{(d-2)}$, i.e.,  a line in $d=3$) \cite{schoendietrich}:
\begin{eqnarray}
\tilde{\Gamma}_\pm(s_\perp,s_\parallel,t;\gamma)&=&
\Gamma^\pm_s(t) s_\perp s_\parallel^{(d-2)} 
+\Gamma^\pm_l(t,\gamma) 
s_\parallel^{(d-2)}\nonumber\\&&+{\cal{O}}(s_\perp^{-1}).
\label{felbontas}
\end{eqnarray}
 While the surface term is determined solely by the order parameter
profile of a semi-infinite binary mixture exposed to a flat substrate, the
line contribution is the specific contribution arising from the change of the
order parameter profile
caused by the edge. Due to the symmetry of the system one can determine
 the amplitudes $\Gamma^\pm_s$ and
$\Gamma^\pm_l$ for the wedge and the ridge explicitly by considering only one 
half of
the wedge or the ridge [see Eq. (\ref{symmetry})], and by suitably 
subtracting and adding
the order parameter profile $m^{\infty/2}_\pm(\zeta=r\sin\theta,t)$ 
of a fluid in contact with an
infinite planar wall in the integrand of Eq.
(\ref{gammahullam}) (compare Eqs. 
(\ref{adszdef})-(\ref{renormaltskala}) and the Appendix): 
\begin{equation}
\Gamma^\pm_s(t)=\Gamma^{\infty/2}_\pm (t)
\label{feluletitag}
\end{equation}
and
\begin{eqnarray}
\Gamma^\pm_l(t,\gamma)&=&2\int_0^{\gamma/2} d\theta 
\int_{0}^\infty dr\: r
[m_\pm (r,\theta,t;\gamma)\nonumber\\
&&-m^{\infty/2}_\pm(\zeta(r,\theta),t)].
\label{redexads}
\end{eqnarray}

Based on Eq. (\ref{skalaekprof}) 
$\Gamma_l^\pm(t,\gamma)$ takes on the scaling form
\begin{equation}
\Gamma_l^\pm(t,\gamma)=a {\xi_0^\pm}^2|t|^{\beta-2\nu}
                             \overline\Gamma_\pm(\gamma)
\label{teljesskala}
\end{equation}
with the universal amplitude functions
\begin{eqnarray}
\overline\Gamma_\pm(\gamma)&=&2\int_0^{\gamma/2} d\theta
\int_{0}^\infty dr_\pm \: r_\pm
[P_\pm(r_\pm,\theta;\gamma)\nonumber \\
&&-P^{\infty/2}_\pm(\zeta_\pm(r_\pm,\theta))]
\label{vegsoskala}
\end{eqnarray}
where $\zeta_\pm(r_\pm,\theta)=r_\pm\sin\theta$. We note that the integral
in Eq. (\ref{vegsoskala}) is finite for $d=4$, i.e.,
$\overline\Gamma_\pm$ does not carry a factor proportional to
$\frac{1}{\nu-\beta}$ as $\Gamma^{\infty/2}_\pm$ does (compare Eq. 
(\ref{renormaltskala})).
As one can see from Eqs. (\ref{renormaltskala}) and (\ref{teljesskala}), the
subdominant line contribution to the excess adsorption carries a more singular
temperature dependence than the leading planar surface term. Thus the
scaling properties of the order parameter profile completely fix the
functional form of the excess critical adsorption up to
the dependence of the universal amplitudes $\overline\Gamma_\pm(\gamma)$ on
the opening angle. Since this dependence cannot be inferred from general
scaling arguments, it must be determined explicitly. This will be carried
out within mean-field theory in the following section. This is possible
because as stated above $\overline\Gamma_\pm(\gamma)$ is finite for $d=4$.


\section{Scaling functions within mean-field theory}

\subsection{Order parameter profiles}

The standard Ginzburg-Landau Hamiltonian for describing critical phenomena
in confined geometries is \cite{binder,diehl}
\begin{equation}
H\{\phi\}=\int_{V_{w(r)}} dV \left\{ \frac{1}{2}(\nabla\phi)^2+\frac{\tau}{2}\phi^2+
\frac{u}{24}\phi^4\right\},
\label{hamiltoni}
\end{equation}
with a scalar order parameter field $\phi({\bm r})$, supplemented by the
boundary condition $\phi=+\infty$ at the surfaces of the wedge (ridge) 
corresponding
to the critical adsorption fixed point \cite{burkdiehl}. The parameter
$\tau$ is proportional to the reduced temperature $t$, $u$ is the coupling
constant, and the integration runs over the volume $V_{w(r)}$ 
of the wedge (ridge) (see Fig. \ref{abra:ek}). Within mean-field theory $\tau=t/(\xi_0^+)^2$ for $T>T_c$, and $\tau=-\frac{1}{2}|t|/(\xi_0^-)^2$ for $T<T_c$ 
with $\xi_0^+/\xi_0^-=\sqrt{2}$.

After functional minimization one obtains for the order parameter
$m=\sqrt{u/12} <\phi>$ the differential equation
\begin{equation}
\Delta m=\tau m +2 m^3,
\label{diffegy}
\end{equation}
where $m=m(r,\theta,\tau;\gamma)$. Since the scaling functions in Eq.
(\ref{skalaforma}) have the limiting behavior shown in Eq.
(\ref{limitbehav}),
 where $\beta=\nu=1/2$ within the mean-field approximation, in
order to derive a boundary condition for the numerical calculation 
 of the order parameter
close to the surfaces of the wedge, i.e., $\theta\rightarrow 0$ for $r$ fixed,
we seek a solution for Eq. (\ref{diffegy}) in the form
\begin{eqnarray}
m(r,\theta,\tau)&=&\frac{A(r,\tau)}{\theta}+B(r,\tau)+C(r,\tau)\theta
+D(r,\tau)\theta^2\nonumber\\
&&+{\cal{O}}(\theta^3)
\end{eqnarray}
for $\theta\ll 1$ (suppressing the $\gamma$-dependence in the notation) 
and obtain for both $\tau>0$ and $\tau<0$
\begin{equation}
m(r,\theta,\tau)=\frac{1}{r}\frac{1}{\theta}+
\left(\frac{1}{6r}-\frac{\tau r}{6}\right)\theta
+{\cal{O}}(\theta^3)
\label{hatar}
\end{equation}
($B(r,\tau)\equiv 0$, $D(r,\tau)\equiv 0$).
This result agrees with the direct expansion of the mean-field profile
obtained for $\tau=0$ \cite{hankestb}.
In terms of the scaling functions $P_\pm$ this implies [see Eq. (\ref{limitbehavek})]:
\begin{subequations}
\begin{equation}
P_+(r_+,\theta\rightarrow 0)=c_+\frac{1}{r_+\sin\theta}
\left[1-\frac{1}{6}r_+^2\theta^2+{\cal{O}}(\theta^4)\right]
\end{equation}
\begin{equation}
P_-(r_-,\theta\rightarrow 0)=c_-\frac{1}{r_-\sin\theta}
\left[1-\frac{1}{12}r_-^2\theta^2+{\cal{O}}(\theta^4)\right]
\end{equation}
\end{subequations}
with $c_+=\sqrt{2}$ and $c_-=2$ \cite{floedie}.

We use a numerical method \cite{colloid,frankdiploma} to minimize Eq.
(\ref{hamiltoni}) with respect to the order parameter profile at a
fixed temperature, which is then subsequently varied. For computational 
purposes we choose suitably shaped finite
volumes $V_{w(r)}$ for different opening angles $\gamma$ of 
the wedge ($w$) or ridge ($r$). 
We refrain from describing this choice of volumes here, because it does not matter in the calculation of the 
universal amplitude functions $\overline\Gamma_\pm(\gamma)$. The
choice of the volume is relevant only for that line contribution to
the excess adsorption that depends only on the order parameter profile
close to an infinite planar wall and thus is independent of the
opening angle of the wedge (see the Appendix).
As the temperature is changed we rescale the size of the volume $V_{w(r)}$
in accordance with the change of the correlation length $\xi=\xi_0 t^{-\nu}$.
This way we control the finite size effects caused by the finiteness of
$V_{w(r)}$ dictated by computational necessity. The finite size effects manifest 
themselves even close to those boundaries of the chosen volume that are far
 away from the
walls of the wedge because of using approximate boundary conditions at these 
boundaries (see below). By effectively increasing the {\em rescaled} volume 
upon approaching $T_c$, the values of the profiles at fixed spatial points 
within this rescaled volume converge to a limiting value.

We choose a two-dimensional grid (i, j) and
calculate the deviation of the order parameter profile from the known
profile at $T=T_c$ at the given grid points. The profile near the confining
surfaces of the wedge (ridge) is fixed according to
Eq. (\ref{hatar}). (The grid points are lined up parallel to the walls
of the wedge.)
At the surfaces of the finite volume $V_{w(r)}$ that are further away from
the walls of the wedge, we prescribe initial values of the 
profile. Keeping these values fixed we then calculate new values of the
profile inside the volume using the method of steepest descent. Having
obtained the new values for the profile close to these surfaces, we change
 the profile at these
surfaces proportionally to the change in their neighborhood, 
if it changes significantly in the direction
perpendicular to the surfaces (the general case), or set it equal to that in neighboring
layers if the profile is approximately constant close to these surfaces
(for example far away from the edges of the wedges parallel to the
confining walls of the wedge).
The rules for the iteration according to the method of steepest
descent in the space of the parameters 
\begin{equation}
a_{ij}=m(r_{ij},\theta_{ij},0)-m(r_{ij},\theta_{ij},\tau)
\end{equation} 
are
\begin{equation}
a_{ij}^{(n+1)}=a_{ij}^{(n)}-\kappa \left.\frac{\partial H({a_{ij}})}{\partial
a_{ij}}\right|_{a_{ij}^{(n)}}
\end{equation}
where 
$\kappa$ is a convergence parameter. With this method it is not necessary
to calculate the Hamiltonian itself but only its derivative. 
The method has been described in detail in
Refs. \cite{colloid} and \cite{frankdiploma}, so here we only want to point 
out that the
divergence of the profile close to the walls of the wedge causes a
divergence in the derivative of the Hamiltonian, too. This can be
avoided, if one takes into account the known form of the divergence of
the profile close to the edges of the wedge [see Eq. (\ref{hatar})].
To this end we write
$a_{ij}$ as a product of two terms, one of which we choose to be
such that when multiplied by the profile close to the boundaries (in
the calculation of the derivative) it cancels
the divergences of the profiles [see Eq. (\ref{hatar})], 
yielding a smooth
gradient in the parameter space. 
We approximate the integrand of the gradient of $H\{\phi\}$ 
[see Eq.({\ref{hamiltoni})]
by a sum of delta-functions positioned at the grid points, so that 
the integral reduces to
a simple sum over these points; for each opening angle 
$\kappa$ is optimized separately for best convergence.

The scaling function of the order parameter profile in a wedge [see Eqs.
(\ref{skalaekprof})-(\ref{symmetry})] with an opening angle of $90^\circ$ is
shown in Figs. \ref{abra:wedgeprofil}-\ref{abra:ivmetszet}.
\begin{figure}[!h]
\includegraphics*[width=\linewidth]{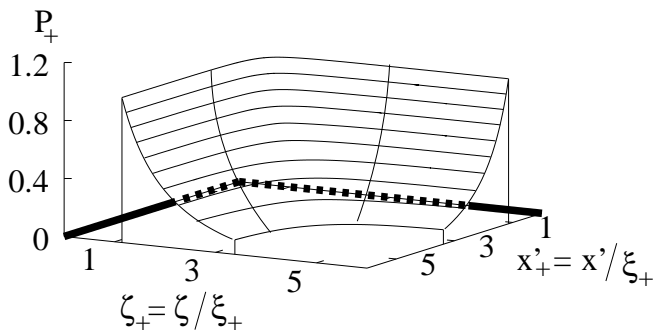}
\caption{The scaling function $P_+$ of the order parameter
profile in a wedge of opening angle $\gamma=\pi/2$ with the
edge located at the back of the figure at $x'=\zeta=0$ 
\{$x'=r\cos\theta$, $\zeta=r\sin \theta$ [see Figs.\ref{abra:ek} and 
\ref{abra:terf2}(a)]\}.
The positions of the walls of the wedge coincide with the coordinate axes 
as indicated here with broad lines.
The values of $P_+$ at which contour lines are drawn are multiples of $0.1$ and range from $0.1$ to $1.0$.}
\label{abra:wedgeprofil}
\end{figure}
One can easily see that the contour lines quickly become parallel to the
walls of the wedge as we move away from the edge
(see Figs. \ref{abra:wedgeprofil} and \ref{abra:2dvonalak}).
 This is
especially apparent as we approach the walls. 
As one moves along the bisector of the wedge, the maximal curvature 
$\kappa$ of the contour lines decreases sharply (Fig. \ref{abra:2dvonalak}).
This underscores that in terms of the {\em rescaled} variables, to a good 
approximation the effects of the edges are spatially localized. 
The maximal curvature $\kappa$ depends linearly on the values of $P_+$
corresponding to the contour lines within the range of $P_+$ values analyzed in 
Fig. \ref{abra:2dvonalak}.
\begin{figure}[!h]
\includegraphics*[width=0.8\linewidth]{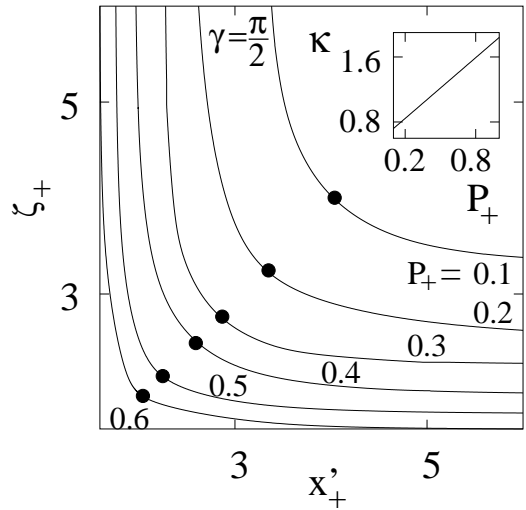}
\caption{Projection of the contour lines of the scaling function $P_+$ of the order parameter
profile in a wedge of opening angle $\gamma=\pi/2$
onto the $(x'_+,\zeta_+)$ plane.
The edge is located at $x'_+=\zeta_+=0$. The curves 
correspond to values of $P_+$ ranging from $0.1$ to $0.6$ with an increment of
$0.1$ from top to bottom. The inset shows that the maximal curvature $\kappa$ 
of the contour lines, occurring on the bisector ($\bullet$), depends linearly 
on the values of $P_+$ corresponding to the contour lines within the range of 
$P_+$ values considered here.
}
\label{abra:2dvonalak}
\end{figure}

Along radial directions, i.e., 
for $\theta=const.$ (Fig. \ref{abra:sugarmetszet}) the scaling function 
exhibits a 
power law limiting behavior close to the walls in accordance with
 Eq. (\ref{limitbehavek}) and the paragraph preceding it. 
For large $r_\pm$ the behavior crosses over into 
an exponential decay:
$P_+(r_+\rightarrow\infty,\theta;\gamma)=A_+(\theta,\gamma)e^{-r_+\sin\theta}$,
$P_-(r_-\rightarrow\infty,\theta;\gamma)-1=A_-(\theta,\gamma)e^{-r_-\sin\theta}$,
where near the walls $A_\pm$ reduce to the amplitudes of the exponential decay
away from an infinite planar wall: $A_+(\theta\rightarrow 0,\gamma)=\sqrt{8}$
and $A_-(\theta\rightarrow 0,\gamma)=2$. The dependence of $A_\pm(\theta,\gamma)$ on 
$\theta$ is weak. The latter values are valid for 
 $\theta\lesssim 30^\circ$, beyond which 
the prefactors of the exponential functions increase with $\theta$.
\begin{figure}[!h]
\includegraphics*[width=\linewidth]{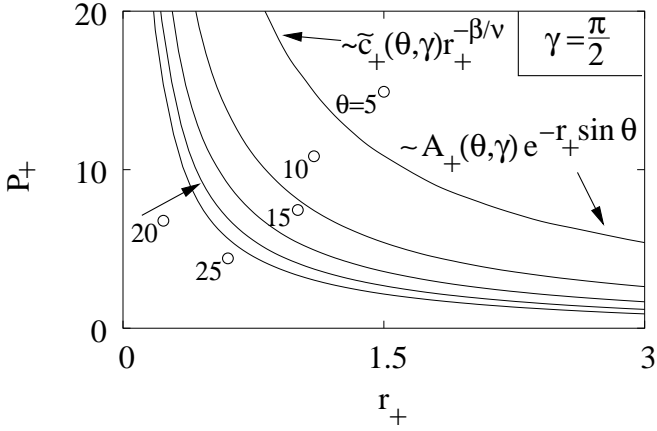}
\caption{The scaling function $P_+$ of the order parameter as a function of the 
distance $r_+$ from the edge of a wedge of opening angle $\gamma=\pi/2$. 
The curves 
correspond to values of $\theta$ ranging from $5^\circ$ to $25^\circ$ with 
an increment of $5^\circ$ from top to bottom.}
\label{abra:sugarmetszet}
\end{figure}
Upon approaching the walls 
of the wedge vertically, i.e., for $\theta\rightarrow 0$ or $\gamma$ 
with $r_\pm$ fixed [see Eq. (\ref{limitbehavek}) and  Fig. \ref{abra:ivmetszet}]
the divergence of the profile has a power law form.
\begin{figure}[!h]
\includegraphics[width=\linewidth]{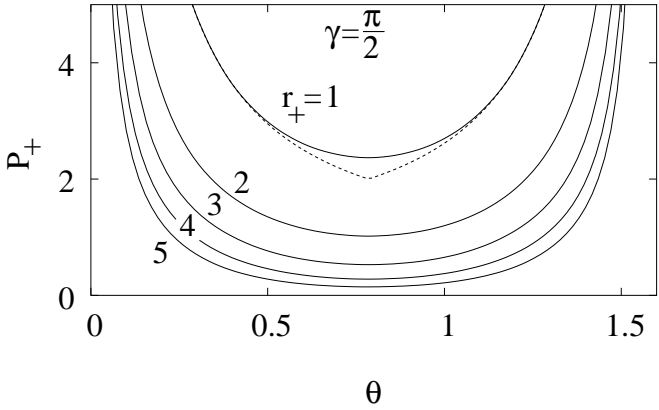}
\caption{The scaling function $P_+$ of the order parameter along arcs of 
different radii $r_+$ centered at the edge of a wedge of opening angle 
$\gamma=\pi/2$. The curves correspond to values of $r_+$ from $1$ to $5$ with 
increments of $1$ from top to bottom. The curves are symmetric around $\theta=\gamma/2=\pi/4$ 
and diverge as $(c_+r_+^{-\beta/\nu})\theta^{-\beta/\nu}$ for $\theta\rightarrow 0$.
For $r_+=1$ the comparison with the asymptotic behavior 
$c_+(r_+\sin\theta)^{-\beta/\nu}$ for $\theta<\gamma/2$ and
$c_+[r_+\sin(\gamma-\theta)]^{-\beta/\nu}$ for $\theta>\gamma/2$ 
is shown as a dashed curve [see Eq. (\ref{limitbehavek})].}
\label{abra:ivmetszet}
\end{figure}

The limiting behaviors of the scaling functions $P_\pm$ close to the edge of the 
wedge ($r_\pm\rightarrow 0$) are described by the amplitude functions 
$\tilde{c}_\pm(\theta,\gamma)$
[see the paragraph following Eq. (\ref{skalaekprof}) 
and Fig. \ref{abra:sugarmetszet}]:
\begin{equation}
P_\pm(r_\pm\rightarrow 0,\theta)=\tilde{c}_\pm(\theta,\gamma)r_\pm^{-\beta/\nu}.
\label{p+-rkicsi}
\end{equation}
These functions are plotted for $\gamma=\pi/2$ in Fig. \ref{abra:chullam}. 
According to Eq. (\ref{limitbehavek}), close to the walls of the 
wedge, i.e., for $\theta\rightarrow 0$ these functions are given by 
$\tilde{c}_\pm(\theta\rightarrow 0,\gamma)=c_\pm(\sin\theta)^{-\beta/\nu}$.
\begin{figure}[!h]
\includegraphics[width=0.8\linewidth]{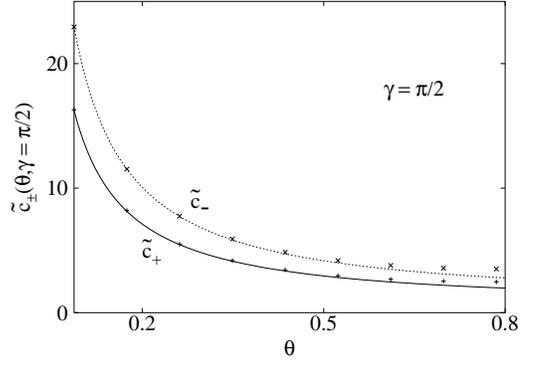}
\caption{The amplitude functions 
$\tilde{c}_\pm(\theta,\gamma=\pi/2)=\left. \left[ P_\pm(r_\pm,\theta)
r_\pm^{\beta/\nu}\right] \right| _{r_\pm\rightarrow 0}$
in a wedge of opening angle $\gamma$ ($t>0 : +$, $t<0 : \times$). The solid line 
corresponds to the function $c_+(\sin\theta)^{-\beta/\nu}$ ($t>0$) [see Eq. 
(\ref{limitbehavek})] and the dotted line to 
$c_-(\sin\theta)^{-\beta/\nu}$ ($t<0$),
 which are valid in the asymptotic regime 
$\theta\rightarrow 0$, but provide a surprisingly good description 
throughout the whole angle range $0<\theta<\pi/4\simeq 0.785$.}
\label{abra:chullam}
\end{figure}

The scaling function of the order parameter profile at a ridge [see Eqs.
(\ref{skalaekprof})-(\ref{symmetry})] with an opening angle of $260^\circ$ is
shown in Figs. \ref{abra:ridgeprofil} and \ref{abra:ridge2dvonalak}.
\begin{figure}[!h]
\includegraphics*[width=\linewidth]{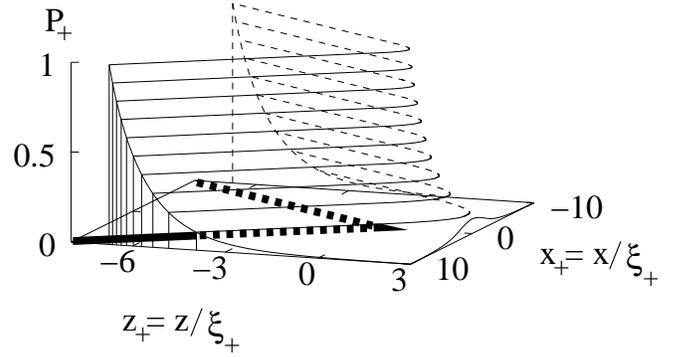}
\caption{The scaling function $P_+$ of the order parameter
profile at a ridge of opening angle $\gamma=260^\circ$ with the
edge perpendicular to the $(x_+,z_+)$ plane and located at $x_+=z_+=0$
 [see Figs. \ref{abra:ek} and, c.f., 
\ref{abra:terf2}(b)].
The positions of the walls of the ridge are indicated here with broad lines.
The values of $P_+$ at which contour lines are drawn are multiples of $0.1$ 
and range from $1.0$ to $0.1$ (top to bottom).}
\label{abra:ridgeprofil}
\end{figure}
One can easily see that the contour lines rapidly become parallel to the
walls of the ridge as one moves along them further away from the edge.
As one moves along the bisector of the ridge away from the edge, the maximal curvature 
$\kappa$ of the contour 
lines decreases (Fig. \ref{abra:ridge2dvonalak}).
\begin{figure}
\includegraphics*[width=0.8\linewidth]{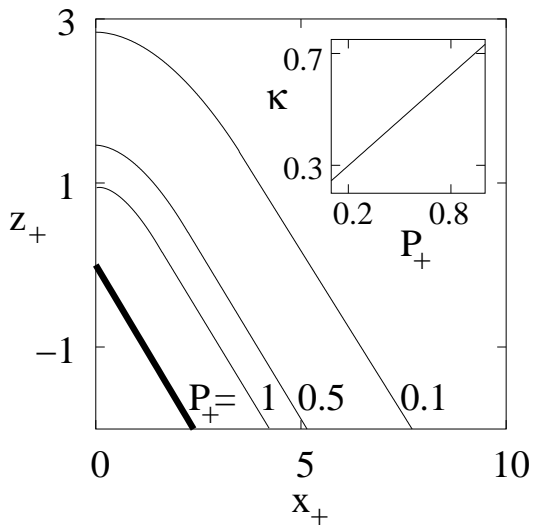}
\caption{Projection of the contour lines of the scaling function $P_+$ of the order parameter profile in a ridge of opening angle $\gamma=260^\circ$ onto the 
$(x_+,z_+)$ plane. The projection is shown only on one side of the bisector 
of the ridge, which coincides here with the 
$z_+$ axis. The wall of the ridge is indicated with a broad line.
The edge is located at $x_+=z_+=0$. The curves 
correspond to $P_+=0.1$, $0.5$, and $1.0$ from top to bottom.  The inset shows that the maximal curvature $\kappa$ 
of the contour lines occurring on the bisector is a linear 
function of $P_+$ within the range considered here.}
\label{abra:ridge2dvonalak}
\end{figure}
 $\kappa$ depends linearly on the values $P_\pm$ characterizing the contour lines
 and thus has similar limiting behaviors close to the edge of the ridge and far from 
it as a function of $r_\pm$, at least within the range studied in the inset of Fig. 
\ref{abra:ridge2dvonalak}. 

Comparing the order parameter profiles in a wedge or at a ridge with the profile 
near a planar wall, one can visualize the wedge or ridge as being formed by breaking the
planar wall into two 
halves, which in the case of a wedge are brought closer to each other, 
and in the case of the ridge are taken further apart. Close to the edge of 
the wedge this increases 
the values of the profiles, while close to the edge of a ridge these values 
are decreased as compared to the profile near a planar wall.  

\subsection{Excess adsorption}

The presentation of the full order parameter distribution requires to keep 
track of four variables: $r,\theta,t$, and $\gamma$. Therefore it is advantageous
to consider also the excess adsorption in wedges and at ridges, which is experimentally 
relevant and provides 
reduced information depending only on $t$ and $\gamma$ 
[Eq. (\ref{gammahullam})]. We are particularly interested in the
line contribution [see Eqs. (\ref{felbontas}) and (\ref{redexads})] 
characterizing the effect of the wedge (ridge) geometry via its universal amplitude
functions [Eq. (\ref{vegsoskala})].
In order to calculate this quantity we use the fact that the integral in 
Eq. (\ref{vegsoskala}) can be rewritten as
\begin{equation}
\overline\Gamma_\pm(\gamma)=\int_0^\infty
F(\overline{P}_\pm) d\overline{P}_\pm,
\label{teruletintegral}
\end{equation}
where $F(\overline{P}_\pm)$ is the area enclosed by the contour lines of
$\overline{P}_\pm=P_\pm(r_\pm,\theta;\gamma)
-P_\pm^{\infty/2}(\zeta_\pm(r_\pm,\theta))$ (see Fig.
\ref{abra:sajatredkontur}).
\begin{figure}
\includegraphics*[width=\linewidth]{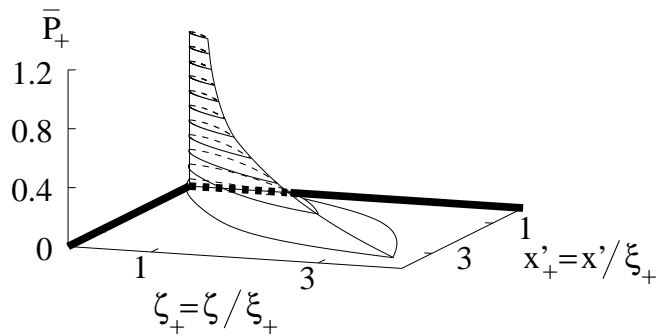}
\caption{Contour levels of the integrand of the integral that determines
the amplitude of the reduced excess adsorption for $T>T_c$
[see Eqs. (\ref{vegsoskala}) and \ref{teruletintegral}]
in a wedge of opening angle $\gamma=\pi/2$ with the
edge located at the back of the figure at $x'=\zeta=0$ ($x'=r\cos\theta$,
$\zeta=r\sin\theta$).
The walls of the wedge, indicated by broad lines, 
 coincide with the coordinate axes.
The values of $\overline{P}_+$ at which contour lines are drawn increase with multiples 
of $0.1$ and range from $0.05$ to $1.05$.}
\label{abra:sajatredkontur}
\end{figure}
Based on these areas one is left with a one-dimensional integration 
to obtain $\overline\Gamma_\pm$ numerically. 
Furthermore, exploiting the
observation that the geometrical shapes formed by the contour lines are
similar to one another for small and large areas, respectively, 
and using the
limiting behavior of the scaling functions, we approximate 
$F(\overline{P}_\pm)$ for small values of $\overline{P}_\pm$ in terms of powers of
$\overline{P}_\pm$, and for large $\overline{P}_\pm$ in term of powers
of $\ln(\overline{P}_\pm)$. These approximate power laws are calculated 
based on different intervals in $\overline{P}_\pm$ chosen as ever narrowing 
slices of that interval in  $\overline{P}_\pm$ in which the numerical data 
lie.
The narrowing intervals approach the small and large $\overline{P}_\pm$ limit, 
respectively.
With these power law approximations for different 
intervals in $\overline{P}_\pm$ we obtain a series of approximate 
integrals for those intervals, for which due to practical limitations there are no
 numerical data (for small and large values of $\overline{P}_\pm$), and take the limit. 
This enables us to carry out the integral in Eq. (\ref{teruletintegral}) numerically 
for the whole range from $\overline{P}_\pm=0$ to $\overline{P}_\pm=\infty$.

This integration leads to the universal amplitudes
$\overline\Gamma_\pm(\gamma)$ as shown in Fig. \ref{abra:adsz}.
\begin{figure}[!h]
\includegraphics*[width=\linewidth]{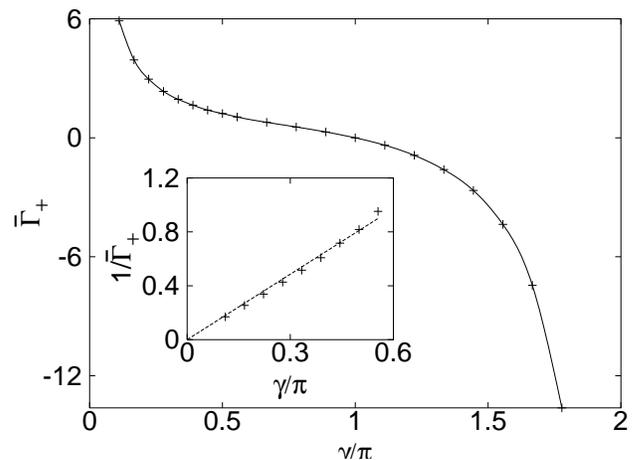}
\caption{The amplitude of the line contribution to the excess adsorption 
for $t>0$ [Eqs. (\ref{teljesskala}) and (\ref{vegsoskala})] as a function of the 
opening angle $\gamma$. The full line is a fit through the data (+).
The inset shows that over a wide range $\overline\Gamma_+(\gamma)\approx
1.62/(\gamma/\pi)$. $\overline\Gamma_+(\pi)=0$ by definition.
Note that $\overline\Gamma_+(\gamma)<|\overline\Gamma_+(2\pi-\gamma)|$.
}
\label{abra:adsz}
\end{figure}
For small opening angles $\gamma$ they diverge as
$\overline\Gamma_\pm(\gamma\rightarrow 0)\sim
1/\gamma$, vary linearly close to  $\gamma=\pi$ for $\gamma\leq\pi$, 
and their absolute values increase rapidly for $\gamma>\pi$. Numerical 
evidence suggests that this latter increase is exponential 
(but no divergence [see Fig. \ref{abra:sorozatgamgam}]).

Strikingly, the reduced excess adsorption $\overline\Gamma_+(\gamma)$ above 
the critical temperature
 appears to be  proportional to the reduced excess adsorption 
$\overline\Gamma_-(\gamma)$ below the critical temperature. We
have calculated their ratio for seven opening angles ranging from $20^\circ$
to $240^\circ$ and found
\begin{equation}
\overline\Gamma_+(\gamma)/\overline\Gamma_-(\gamma)=1.137\pm 0.006,
\label{arany}
\end{equation}
i.e., their ratio appears to be independent of $\gamma$.

Ratios of the amplitudes of the excess adsorption above and below $T_c$ 
have been investigated for the case of a planar wall theoretically 
\cite{floedie},
experimentally \cite{smithlaw}, and using Monte Carlo simulations
\cite{sdl}. 
The values obtained experimentally for the ratios of the
amplitudes for the planar case (mean value: $1.19\pm 0.04$) agree rather well
 with the result of
the Monte Carlo simulations ($1.11$); the corresponding mean field value is 
$1/\sqrt{2}$ \cite{floedie}.

The angular dependence of the reduced excess adsorption shows that
for large angles, i.e., for a ridge, the absolute values are larger 
than for small angles, i.e., for a wedge:
$\overline\Gamma_\pm(\gamma)<|\overline\Gamma_\pm(2\pi-\gamma)|$. 
One can calculate the excess adsorption at a periodic array of wedges and ridges 
(see, c.f., Fig. \ref{abra:sok})
if their edges are sufficiently far apart from each other so that their influences 
do not interfere (see part B of the Appendix). The result expressed by Eq. 
\ref{sorozatfelb} in the Appendix shows that in this limiting case the line 
contribution to the excess 
adsorption, which captures the effect of the wedges and ridges, is the sum of 
the contributions of single wedges of opening angle $\gamma_w$ and of 
single ridges of
opening angle $\gamma_r=2\pi -\gamma_w$ sharing the same temperature dependence 
$|t|^{\beta-2\nu}$. Thus the amplitude $\overline\Gamma_\pm^{wr}$ 
of the combined contribution of one wedge and a neighboring ridge forming the basic 
building block of the array is given as  
as $\overline\Gamma_\pm^{wr}=\overline\Gamma_\pm(\gamma_w)+\overline\Gamma_\pm
(\gamma_r)$. This quantity may be viewed as a function of 
$\gamma_r-\gamma_w=2\pi -2\gamma_w=
2(\pi-\gamma_w)$, which characterizes the roughness of the surface.
$\overline\Gamma_+^{wr}(\gamma_r-\gamma_w)$ is plotted in Fig. 
\ref{abra:sorozatgamgam} for ($t>0$). One can see that all values are negative, i.e.,
the total excess adsorption (relative to a planar substrate with the same area as the 
actual one of the corrugated surface) is decreased by the line contribution.
This demonstrates that the decrease in adsorption for a ridge with 
$\gamma_r=2\pi -\gamma_w$ dominates the increase due to the corresponding wedge with 
opening angle $\gamma_w$ (see Fig. \ref{abra:adsz}).
The amplitude $\overline\Gamma_+^{wr}$ varies quadratically for small roughness and 
exponentially for large roughness.
\begin{figure}[!h]
\includegraphics*[width=\linewidth]{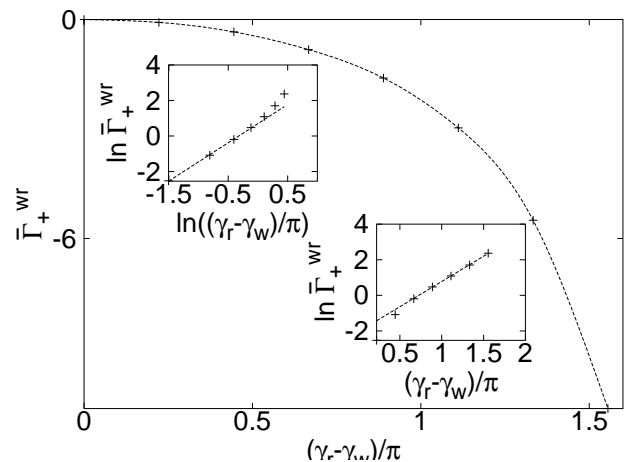}
\caption{The amplitude $\overline\Gamma_+^{wr}$ of the line contribution 
to the excess adsorption of the basic unit of one wedge of opening angle $\gamma_w$ 
and one ridge of opening angle $\gamma_r=2\pi -\gamma_w$ in an array of effectively 
independent wedges and ridges as a function of $(\gamma_r-\gamma_w)/\pi$, 
which characterizes the roughness of the surface (see, c.f., Fig. \ref{abra:sok}). 
The dashed lines are fits.
 The insets show the behaviors in the limiting cases on a log-log
scale and a log-linear scale for small and large roughness, respectively.
They indicate that $\overline\Gamma_+^{wr}((\gamma_r-\gamma_w)\rightarrow 0)
\sim(\gamma_r-\gamma_w)^2$ and that $\overline\Gamma_+^{wr}$ increases as 
$exp\{[(\gamma_r-\gamma_w)/\pi]^3\}$ up to $(\gamma_r-\gamma_w)/\pi=2$.}
\label{abra:sorozatgamgam}
\end{figure}

In Ref. \cite{hankedietrich} 
the total excess adsorption has been calculated for curved surfaces. 
For a curved membrane with both sides exposed to a
fluid near criticality, the sum of the excess adsorptions on the two sides
per unit area was found to be larger for spherical regions of the membrane
and smaller for cylindrical regions as compared to that for flat regions.
Since the cylindrical regions may be viewed as rounded wedges and
ridges, this latter finding exhibits the same qualitative trend as found here for the 
periodic array of wedges and ridges.


\section{Free energy of the confined fluid and Casimir torque}
\label{szabadenergfej}

In the previous chapters we have investigated structural properties,
 i.e., order parameter profiles of critical fluids
 in wedges and close to ridges. 
 In the followings we comment on their thermodynamic properties based on
the free energy of such systems. 

The volume $V(\gamma)$ of the system shown in, c.f.,
Fig. \ref{abra:terf1} 
(for another
possible choice see, c.f., Fig. \ref{abra:terf2}) is
bounded by the walls of the wedge or ridge, which in turn end in the linear
edge forming the vertex. Accordingly in the thermodynamic limit the
total free energy F
decomposes into a bulk, a
surface, and a line contribution:  
\begin{equation}
F(T;\gamma)=V(\gamma)f_b(T)+Sf_s(T)+Lf_l(T,\gamma),
\end{equation}
where $f_b(T)$ is the bulk free energy density, $f_s(T)$ is the surface free
energy density, and $f_l(T,\gamma)$ is the line free energy density. $S$ is
the total surface area of the wall in contact with the fluid, $L$ is the 
length of the
edge. Each of the three terms in the free energy and thus the total free energy itself
are sums of a singular part ($f_{sing}^{(b,s,l)}(t,\gamma)$), which contains the
thermodynamic singularities  in the vicinity
of the bulk critical point $t=(T-T_c)/T_c\rightarrow 0$, and an analytic 
background contribution 
($f_{back}^{(b,s,l)}(T,\gamma)$).

The leading singular part of the total free energy can be written in the 
form (see, e.g., Refs. \cite{diehl}, \cite{dietrichdiehl}, and \cite{privmanhoheaha}):
\begin{multline}
\frac{F_{sing}^\pm(t,\gamma)}{k_BT_c}=\frac{V(\gamma)}{(\xi_0^\pm)^d}
\left(-\frac{a_b^\pm}{\alpha(1-\alpha)(2-\alpha)}|t|^{2-\alpha}\right)
\\
+\frac{S}{(\xi_0^\pm)^{d-1}}
\left(-\frac{a_s^\pm}{\alpha_s(1-\alpha_s)(2-\alpha_s)}|t|^{2-\alpha_s}\right)
\\
+\frac{L}{(\xi_0^\pm)^{d-2}}
\left(-\frac{a_l^\pm(\gamma)}{\alpha_l(1-\alpha_l)(2-\alpha_l)}|t|
^{2-\alpha_l}\right).
\label{szabenergskala}
\end{multline}
Here $\alpha\simeq0.11$ is
the bulk specific heat exponent, $\alpha_s=\alpha+\nu$,
$\alpha_l=\alpha+2\nu$, and $a_b^\pm$ and $a_s^\pm$ are universal bulk and
surface amplitudes. The bulk contribution depends trivially on $\gamma$ via
the geometry $V(\gamma)$, whereas the surface contribution is independent of
$\gamma$. The line contribution carries a nontrivial dependence on $\gamma$
via the universal amplitude functions $a_l^\pm(\gamma)$.

The background contribution takes on the form
\begin{eqnarray}
\frac{F_{back}^\pm(T,\gamma)}{k_BT_c}&=&\frac{V(\gamma)}{(\xi_0^\pm)^d}
f_{back}^{(b)}(T)
+\frac{S}{(\xi_0^\pm)^{d-1}}
f_{back}^{(s)}(T)\nonumber\\
&&+\frac{L}{(\xi_0^\pm)^{d-2}}
f_{back}^{(l)}(T,\gamma).
\label{szabadback}
\end{eqnarray}
If one of the sidewalls is moveable around the vertex with the far end
suspended at, say, a force microscope, the torque
\begin{equation}
M=-\frac{\partial{F(T;\gamma)}}{\partial{\gamma}}
\label{forgnyomdef}
\end{equation}
exerted by the fluid in the wedge or ridge on its sidewalls is experimentally 
accessible:
\begin{widetext}
\begin{eqnarray}
\frac{M^\pm}{k_BT_c}&=&-\frac{\partial V(\gamma)}{\partial \gamma}(\xi_0^\pm)^{-d}
\left(-\frac{a_b^\pm}{\alpha(1-\alpha)(2-\alpha)}|t|^{2-\alpha}
+f_{back}^{(b)}(T)\right)
\nonumber\\
&&-L(\xi_0^\pm)^{-(d-2)}
\left(-\frac{\partial a_l^\pm(\gamma)/\partial \gamma}
{\alpha_l(1-\alpha_l)(2-\alpha_l)}|t|
^{2-\alpha_l}
+\frac{\partial f_{back}^{(l)}(T,\gamma)}{\partial \gamma}
\right).
\label{forgnyom}
\end{eqnarray}
\end{widetext}
With the bulk contribution known independently, this measurement provides
access to the universal amplitude functions $a_l^\pm(\gamma)$ by focusing on
the thermal singularity $\sim |t|^{2-\alpha_l}= |t|^\nu = |t|^{0.63}$
in $d=3$, since $d\nu=2-\alpha$.
The singular contribution to $M$ can be called a critical Casimir torque. For
fluids information about the background term can be obtained from Eq. (A7) in
Ref. \cite{gettadietrich} and from Refs. \cite{hendphysa, hendjchem, hendprep}.

Within mean-field theory $2-\alpha_l=1$ so that the singular line contribution
becomes indistinguishable from the analytical background contribution. Thus
our present approach renders only the sum of these two contributions without
the possibility to isolate the amplitude functions $a_l^\pm(\gamma)$. As
indicated by the pole $\sim \frac{1}{1-\alpha_l}$ for $d\rightarrow 4$,
inclusion of Gaussian fluctuations beyond the simple mean-field theory is
expected to generate a term $\sim t\ln |t|$ due to the resonance of the
singular contribution $\sim \frac{1}{1-\alpha_l}t^{2-\alpha_l}$ with an
analytical background term $\sim t$ \cite{privman}. The amplitude of the
singular term $\sim t\ln |t|$ would allow one to retrieve at least the
mean-field expression for $a_l^\pm(\gamma)$. However, this technically
challenging inclusion of Gaussian fluctuations is beyond the scope of the
present work.

A suitable approach to obtain the change of the free energy upon varying the 
opening angle of the wedge or ridge involves
calculating the field theoretical stress tensor. 
Analogously to the free energy density,
the corresponding torque requires additive renormalization 
 up to second order in temperature \cite{krechdierenorm}. We have followed
this route within the present mean-field theory without isolating the critical 
Casimir torque. We have found that this combined torque
diverges as $M\sim 1/\gamma^2$ for small $\gamma$, and it appears to be a linear 
function of $1/\gamma^{2}$ within the angle range between $\gamma=20^\circ$
and $\gamma=280^\circ$.


\section{Summary}

In the present study of critical adsorption in wedges and close to ridges (see
Fig. \ref{abra:ek}) we have obtained
the following main results:

(1) We have discussed the scaling properties of the order parameter profile
    $m_\pm(\bm{r},t;\gamma)=a|t|^\beta P_\pm(r/\xi_\pm,\theta;\gamma)$
    in terms of the bulk correlation length $\xi_\pm=\xi_0^\pm
    |t|^{-\nu}$ above and below the critical point $T_c$ with $t=(T-T_c)/T_c$. 
 The universal scaling functions
    $P_\pm(r/\xi_\pm,\theta;\gamma)$ diverge according to a power law close to
    the walls of the wedge or ridge, and decay exponentially far away from the
    walls [Eq. (\ref{limitbehavek})].

(2) In the thermodynamic limit the excess adsorption
    $\tilde{\Gamma}_\pm(s_\perp,s_\parallel,t;\gamma)$ [Eq. (\ref{gammahullam})]
    for volumes with linear extensions $s_\perp$ and $s_\parallel$ 
(see Figs. \ref{abra:terf1} and \ref{abra:terf2})
 decomposes into a surface contribution that scales with the
actual surface area of the confining walls
($s_\perp s_\parallel^{(d-2)}$) and a line
contribution that scales with the extension in the invariant directions
($s_\parallel^{(d-2)}$) [see Eq. (\ref{felbontas})] as described in
    detail in the
    Appendix. The line contribution is the specific contribution arising
    from the influence of the edge on the order parameter profile. Its
    amplitude $\Gamma^\pm_l(t,\gamma)$ has the scaling form
    $\Gamma_l^\pm(t,\gamma)=
a {\xi_0^\pm}^2|t|^{\beta-2\nu} \overline\Gamma_\pm(\gamma)$
with the universal amplitude functions $\overline\Gamma_\pm(\gamma)$
    [Eq. (\ref{vegsoskala})] carrying
    the dependence on the opening angle $\gamma$.

(3) We calculate the above scaling functions within mean-field theory [see
    Eqs. (\ref{hamiltoni}) and (\ref{diffegy})] using a numerical algorithm both
    above and below the critical temperature (for a
    wedge see Figs. \ref{abra:wedgeprofil}-\ref{abra:ivmetszet}, for a
    ridge see 
Figs. \ref{abra:ridgeprofil} and \ref{abra:ridge2dvonalak}). The amplitude
    functions of the power law divergence of the profile close to the walls
    [Eq. (\ref{limitbehavek})] are shown in Fig. \ref{abra:chullam}.

(4) Our numerical calculation also yields the experimentally relevant
    excess adsorption
    within mean-field approximation. The universal amplitudes
$\overline\Gamma_+(\gamma)$ are shown in Fig. \ref{abra:adsz}.
 For small opening angles $\gamma$ they diverge as
$\overline\Gamma_+(\gamma\rightarrow 0)\sim
1/\gamma$, vary linearly close to  $\gamma=\pi$ for $\gamma\leq\pi$,
and their absolute values increase rapidly for $\gamma>\pi$. Numerical
evidence suggests that this latter increase is exponential,
but without divergence [see Fig. \ref{abra:sorozatgamgam}].
The reduced excess adsorption $\overline\Gamma_+(\gamma)$ above
the critical temperature
 appears to be  proportional to the reduced excess adsorption
$\overline\Gamma_-(\gamma)$ below the critical temperature with 
$\overline\Gamma_+(\gamma)/\overline\Gamma_-(\gamma)=1.137\pm 0.006$.
We have considered a wedge and a ridge together as forming the basic unit in a
    periodic array (see Fig. \ref{abra:sok}). The total excess adsorption
    relative to that of a planar substrate with the same area as the
actual one of the corrugated surface [Eq. (\ref{sorozatextra})] is decreased by 
the line contribution (see Fig. \ref{abra:sorozatgamgam}).

(5) The variation of the free energy of the system with the opening
angle of the wedge or ridge gives rise to a torque acting on the sidewalls
[Eq. (\ref{forgnyomdef})]. The free
    energy decomposes into a singular contribution exhibiting scaling
    [see Eq. (\ref{szabenergskala})], and an analytic background contribution
    [Eq. (\ref{szabadback})]. 
In $d=3$ the critical Casimir torque varies as
$a_l^\pm(\gamma)|t|^{\nu}$ with universal amplitude functions
$a_l^\pm(\gamma)$. This cusplike temperature singularity is expected
to be experimentally accessible via suitable force microscopy. The
theoretical calculation of the corresponding amplitude functions
remains as a challenge.


\begin{acknowledgments} 
G. P. is indebted to F. Schlesener and M. Krech for
many helpful discussions that settled numerous technical and physical questions.
\end{acknowledgments}


\appendix*
\section{Decomposition of the excess adsorption}
\label{decompapp}

In this appendix we discuss how the excess adsorption in a wedge or at a
ridge decomposes into surface and line contributions [see Eqs.
(\ref{felbontas})-(\ref{redexads})].
In the definition of the excess adsorption [Eq. (\ref{gammahullam})] one
considers a finite volume $V$ of integration that is enlarged to fill the
total volume of the wedge or ridge in the thermodynamic limit. As shown
 below for two examples, the expression for the line term (such as Eq.
(\ref{redexads})) actually depends on the choice
of the shape of the volume $V$.

In the following we first analyze the excess adsorption for a single
macroscopic wedge
 or ridge (\ref{egywedgeridge}), which will be followed by a discussion of
the excess adsorption for two possible experimental realizations
(\ref{sorozatapp} and \ref{egywedgeapp}).

\subsection{A single macroscopic wedge or ridge}
\label{egywedgeridge}

\subsubsection{First choice of the volume of integration}
\label{subsub1}

Our first choice for $V$ is shown in Fig. \ref{abra:terf1} for the case
of a wedge. This choice is inspired by the idea that the single wedge or
ridge
considered here is ultimately a member of a periodic array.
 With this polygonal cross section
and a similarly constructed one for the
ridge, one can cover the total volume $V_{tot}$ 
of a fluid in contact with a surface formed as a periodic array
 of wedges and ridges in a natural way (see Fig. \ref{abra:sok}). 
All the formulas explicitly stated below for the wedge are valid for the
ridge, too.
\begin{figure}[!h]
\includegraphics*[width=0.8\linewidth]{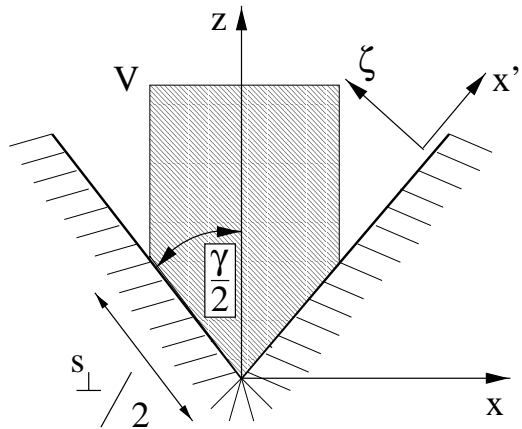}
\caption{One choice for the volume $V$ of integration in the
definition of the excess adsorption [Eq. (\ref{gammahullam})] with its 
cross section
in the plane perpendicular to the invariant directions. }
\label{abra:terf1}
\end{figure}
\begin{figure}[!h]
\includegraphics*[width=\linewidth]{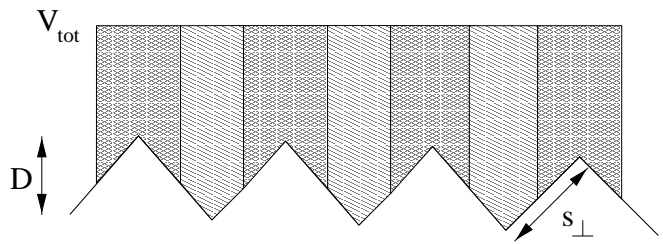}
\caption{A fluid is bounded by a substrate shaped as a periodic array of
wedges and ridges with a finite depth $D$.
The total volume of the fluid $V_{tot}$ can be naturally decomposed into 
subvolumes of the type shown in Fig. \ref{abra:terf1}.}
\label{abra:sok}
\end{figure}
The volume $V$ 
is symmetric with respect to the bisector plane of the wedge, which allows
us in the following to consider only one half of the wedge, and multiply the
corresponding expression by a factor of two.

As indicated in Fig. \ref{abra:terf1} the volume $V$ in Eq.
(\ref{gammahullam}) is finite and thus the integral 
is finite, too. As a first step in carrying out the thermodynamic limit we
keep the shape of $V$ but shift the upper boundary $z=const$ to infinity.
Since $m_\pm$ approaches $m_b$ exponentially, this extension of $V$
increases $\tilde{\Gamma}_\pm$ by an exponentially small amount and thus
does not contribute to the two leading terms under consideration in Eq.
(\ref{felbontas}). (In the spirit of the thermodynamic limit one first
increases $V$ before one can possibly consider the limit $t\rightarrow 0$.
Therefore these arguments are not impaired by a power law decay of
$m_\pm(z\rightarrow \infty, t=0)$ in the thermodynamic limit.)
 For the resulting semi-infinite strip Eq.
(\ref{gammahullam}) can be rewritten in the following form 
by adding and subtracting the order
parameter profile $m^{\infty/2}_\pm(\zeta=r\sin\theta,t)$ 
of a fluid in contact with an
infinite planar wall in the integrand (see Figs.
\ref{abra:ek} and \ref{abra:terf1}):
\begin{widetext}
\begin{eqnarray}
\tilde{\Gamma}_\pm(s_\perp,s_\parallel,t;\gamma)&=&
2 s_\parallel^{d-2}\int_{0}^\infty d\zeta 
\int_{\frac{\zeta}{\tan(\gamma/2)}}
^{\frac{s_\perp}{2}+\frac{\zeta}{\tan(\gamma/2)}} dx'
[m_\pm(r(x',\zeta),\theta(x',\zeta),t;\gamma)-m^{\infty/2}_\pm(\zeta,t)]
\nonumber\\&&+
2 s_\parallel^{d-2}\int_{0}^\infty d\zeta 
\int_{\frac{\zeta}{\tan(\gamma/2)}}
^{\frac{s_\perp}{2}+\frac{\zeta}{\tan(\gamma/2)}} dx'
[m^{\infty/2}_\pm(\zeta,t)-m_b(t)]
\label{hosszugammahullam}
\end{eqnarray}
\end{widetext}
where $x'=r\cos\theta=z\cos(\gamma/2)+x\sin(\gamma/2)$ is 
the coordinate measured from the apex parallel 
to the nearest wall of the wedge and 
$\zeta=r\sin\theta=z\sin(\gamma/2)-x\cos(\gamma/2)$ is 
the normal distance from the nearest wall of the wedge (see Fig.
\ref{abra:terf1}). 

In the inner integral of the first term the upper integration limit can be 
shifted to infinity, i.e., $s_\perp\rightarrow +\infty$, with an addition of
only exponentially small corrections to the integral, 
because $m_\pm$ approaches
$m_\pm^{\infty/2}$ exponentially for $x'\rightarrow +\infty$ at fixed 
$\zeta$. Thus the first term in Eq. (\ref{hosszugammahullam}) approaches a
constant for $s_\perp\rightarrow +\infty$ and this constant involves an
unlimited integral over the whole half of the wedge.
Expressed in terms of cylindrical coordinates this term yields the
line contribution 
$\Gamma^\pm_l(t,\gamma)s_\parallel^{(d-2)}$ in Eqs. (\ref{felbontas}) and
(\ref{redexads}).

As the integrand of the second term in Eq. (\ref{hosszugammahullam}) 
does not depend on $x'$, the inner integration simply yields a factor
$s_\perp/2$. The outer integral then yields $\Gamma_\pm^{\infty/2}$ in Eq.
(\ref{adszdef}). Together with Eq. (\ref{feluletitag}) this verifies Eq.
(\ref{felbontas}).


\subsubsection{Second choice of the volume of integration}
\label{subsub2}

Our second choice of the shape of the volume $V$ as shown in Fig.
\ref{abra:terf2}(a) corresponds to the one used in Ref.
\cite{schoendietrich}, where liquids confined by two opposing structured
walls have been studied; in this geometry one cannot infinitely extend 
the volume in the $+z$ direction. On the other hand in the case of the
ridge as shown in Fig. \ref{abra:terf2}(b),
\begin{figure}[!h]
\includegraphics*[width=0.65\linewidth]{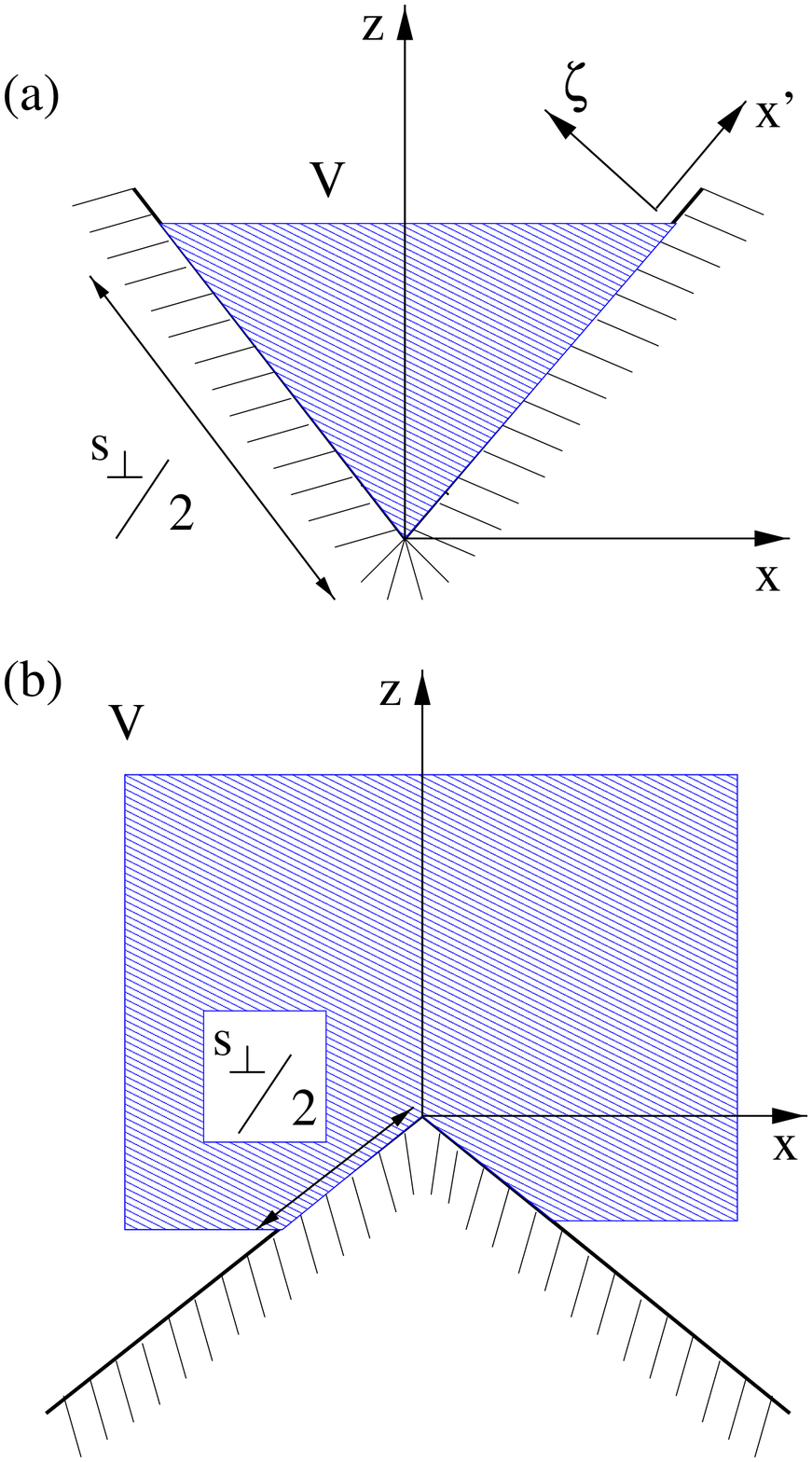}
\caption{A second choice for the volume $V$ of integration in the
definition of the excess adsorption [Eq. (\ref{gammahullam})] for a wedge
(a) and a ridge (b) with its cross section
in the plane perpendicular to the invariant directions. }
\label{abra:terf2}
\end{figure}
starting out from a finite volume $V$, the upper boundary of the volume
($z=const$) can be shifted to $+\infty$ as well as the two vertical boundaries
($x=\pm const$) to $\pm\infty$ with only exponentially small corrections
to the integral in Eq. (\ref{gammahullam}), because in the directions
$z\rightarrow\infty$ and $x\rightarrow\pm\infty$ for $z$ fixed the order
parameter attains its bulk value exponentially.
As for the previous choice also for the present geometries Eq.
(\ref{gammahullam}) can be rewritten by adding and subtracting the order
parameter profile $m^{\infty/2}_\pm(\zeta=r\sin\theta,t)$ 
of a fluid in contact with an
infinite planar wall in the integrand (see Fig.
\ref{abra:terf2}):
\begin{widetext}
\begin{eqnarray}
\tilde{\Gamma}_\pm(s_\perp,s_\parallel,t;\gamma)&=&
2 s_\parallel^{d-2}\int_{0}^{\frac{s_\perp}{4}\sin\gamma} d\zeta 
\int_{\frac{\zeta}{\tan(\gamma/2)}}
^{\frac{s_\perp}{2}-\zeta\tan(\gamma/2)} dx'
[m_\pm(r(x',\zeta),\theta(x',\zeta),t;\gamma)-m^{\infty/2}_\pm(\zeta,t)]
\nonumber\\&&+
2 s_\parallel^{d-2}\int_{0}^{\frac{s_\perp}{4}\sin\gamma} d\zeta 
\int_{\frac{\zeta}{\tan(\gamma/2)}}
^{\frac{s_\perp}{2}-\zeta\tan(\gamma/2)} dx'
[m^{\infty/2}_\pm(\zeta,t)-m_b(t)]
\end{eqnarray}
\end{widetext}
where as before $x'=r\cos\theta$ is the coordinate parallel 
to the wall, and $\zeta=r\sin\theta$ normal to it [see Fig.
\ref{abra:terf2}(a)]. 
In the first term the upper integration limits of both integrals can be
shifted to infinity, i.e., $s_\perp\rightarrow\infty$, with an addition of
only exponentially small corrections to the integral, because $m_\pm$ 
approaches $m_\pm^{\infty/2}$ exponentially for $x'\rightarrow +\infty$ at 
fixed $\zeta$, and $m_\pm$ attains its bulk value exponentially 
for $\zeta\rightarrow +\infty$ at fixed $x'$. Thus for the first term the
limit $s_\perp\rightarrow\infty$ exists and is finite with the
two-dimensional integral covering the whole half of the wedge. Expressed in
terms of cylindrical coordinates this term yields the line term 
$\Gamma^\pm_l(t,\gamma)s_\parallel^{(d-2)}$ [see Eqs. (\ref{felbontas}) and
(\ref{redexads})] as for the previous choice of the volume. 

The inner integral in the second term can be carried out, because the
integrand is independent of $x'$. This yields a prefactor 
$\frac{s_\perp}{2}-\zeta\left[\tan\frac{\gamma}{2}+\frac{1}{\tan\gamma/2}\right]=
\frac{1}{2}\left\{s_\perp-\frac{4}{\sin\gamma}\zeta\right\}$.
The first term of this prefactor gives rise to the surface contribution in Eqs.
(\ref{adszdef}), (\ref{felbontas}), and (\ref{feluletitag}), if one shifts
the upper integration limit of the $\zeta$ integration to infinity with the
addition of exponentially small corrections. After multiplying this
prefactor by two, its second term gives rise, however, to another line 
contribution with the amplitude
\begin{equation}
\hat{\Gamma}^\pm_l(t,\gamma)=\frac{-4}{\sin\gamma}
\int_0^\infty \zeta[m^{\infty/2}_\pm(\zeta,t)-m_b(t)] d\zeta,
\label{extraterm}
\end{equation}
where the upper limit of integration has also been shifted to infinity 
with an exponentially small correction. 
We note that this additional line term depends 
on the order parameter profile at a planar substrate only. 
Due to the extra factor $\zeta$ in the integrand,
the integral in Eq. (\ref{extraterm}) is finite for $d=4$
in spite of $m_\pm^{\infty/2}(\zeta\rightarrow 0)\sim\zeta^{-1}$ in this case, 
i.e., $\hat{\Gamma}_\pm$ does not carry a factor proportional to
$\frac{1}{\nu-\beta}$ as $\Gamma^{\infty/2}_\pm$ does (compare Eq. 
(\ref{renormaltskala})). 

Thus in the thermodynamic limit the two choices of the integration volume
$V$ yield the same surface contributions $\Gamma_s^\pm$ to the excess adsorption
but different subdominant line contributions $\Gamma_l^\pm$:
\begin{subequations}
\begin{equation}
\int_V d^dx\:m_\pm(\bm{x})=Vm_b+S\Gamma_s^\pm+L\Gamma_l^\pm+{\cal{O}}(s_\perp^{-1})
\label{altfelbontas}
\end{equation}
\text{with}
\begin{eqnarray}
\Gamma_l^\pm&=&\Gamma_{l,I}^\pm(t,\gamma)=
\nonumber\\&&2\int_0^{\gamma/2} d\theta 
\int_{0}^\infty dr\: r
[m_\pm(r,\theta,t;\gamma)
\nonumber\\&&-m^{\infty/2}_\pm(\zeta(r,\theta),t)],\quad
\text{choice I,}
\label{altfel1terf}
\end{eqnarray}
\text{and}\\
\begin{equation}
\Gamma_l^\pm=\Gamma_{l,II}^\pm(t,\gamma)=\Gamma_{l,I}^\pm(t,\gamma)+
\hat{\Gamma}_l^\pm(t,\gamma),\quad\text{choice II,}
\label{altfel2terf}
\end{equation}
\end{subequations}
with $S=s_\perp s_\parallel^{d-2}$, $L=s_\parallel^{d-2}$, and
$\hat{\Gamma}_l^\pm$ given by Eq. (\ref{extraterm}). Experiments
cannot be carried out for infinitely deep wedges. Instead they can be
carried out for either a periodic array of wedges or for a single wedge of
finite depth carved out from a wide planar surface. In both cases additional
ridges must be formed giving rise to their own adsorption properties.
Therefore experiments on such systems give access only to certain
combinations of wedge and ridge excess adsorptions, whose corresponding line
contributions carry the relevant additional information about the nonplanar
substrate geometry. The actual choice of the corresponding integration
volume depends on the actual experimental setup (compare, e.g., Fig.
\ref{abra:sok}; see Subsecs. \ref{sorozatapp} and \ref{egywedgeapp})
The results of this subsection show that the line
contributions depend on such details even in the thermodynamic limit.


\subsection{Periodic array of wedges and ridges}
\label{sorozatapp}

In this subsection we consider a substrate with a periodic series of edges 
and wedges as depicted in
Fig. \ref{abra:sok}. 
There is a variety of experimental techniques to create such kind of surface
morphology.
If the opening angle of the wedges is $\gamma$, 
the opening angle of the ridges is $2\pi-\gamma$. Here we focus on the
limiting case that the depth of the wedges
$D=s_\perp\cos(\gamma/2)$ 
is sufficiently large, so that the deviations of the profiles close to the
edges of the wedges and ridges, respectively, from the profile of a fluid
 exposed to an infinite
planar substrate do not influence each other. As apparent from Figs.
\ref{abra:terf1} and \ref{abra:sok}, this case corresponds to the first
choice of the volume of integration for the single wedge or ridge and 
thus leads to the
following decomposition of the
excess adsorption [see Eq. (\ref{altfel1terf})]
\begin{eqnarray}
&&\int_{V_{tot}}
d^dx\:m_\pm(\bm{x},t;\gamma)=
V_{tot}m_b(t)+S_{tot}\Gamma_s^\pm (t)
\nonumber\\&&+L(N_w\Gamma_{lw}^\pm(t,\gamma_w)
+N_r\Gamma_{lr}^\pm(t,\gamma_r))+
{\cal{O}}(s_\perp^{-1})
\label{sorozatfelb}
\end{eqnarray}
with the total surface of the substrate $S_{tot}=Ns_\perp s_\parallel^{d-2}$, 
where $N=N_w+N_r$ is the number of segments of length $s_\perp$,
$\Gamma^\pm_s(t)=\Gamma^{\infty/2}_\pm (t)$ [see Eq. (\ref{adszdef})],
$L=s_\parallel^{d-2}$, $N_w=N_r$ are the numbers of wedges and ridges,
respectively, and
\begin{widetext}
\begin{equation}
\Gamma^\pm_{lw(r)}(t,\gamma_{w(r)})=
2\int_0^{\gamma_{w(r)}/2} d\theta 
\int_{0}^\infty dr\: r
[m^{w(r)}_\pm(r,\theta,t;\gamma_{w(r)})
-m^{\infty/2}_\pm(\zeta(r,\theta),t)],
\label{sorozatextra}
\end{equation}
\end{widetext}
where $\gamma_w$ is the opening angle of the wedge, and 
$\gamma_r=2\pi-\gamma_w$.
This is in accordance with Eqs. (32)-(36) of Ref. \cite{schoendietrich}, where
a different coordinate system was used \footnote{In obtaining the first term
in Eq. (34) of Ref. \cite{schoendietrich}, a different scaling of the profile
of a fluid exposed to 
 an infinite planar substrate was used as a reference for the wedge and the 
ridge part, which caused this extra term to appear. If the same scaling is
used, as it is correct, this first term becomes zero in accordance with the
present findings in Eqs. (\ref{sorozatfelb}) and (\ref{sorozatextra}).}.


\subsection{A single wedge embedded into a planar wall}
\label{egywedgeapp}

In this subsection we consider a single wedge of opening angle $\gamma$ 
carved out of a planar surface, thus producing also two ridges of opening
angles $(3\pi-\gamma)/2$ (see Fig. \ref{abra:ekfal}). 
\begin{figure}[!h]
\includegraphics*[width=\linewidth]{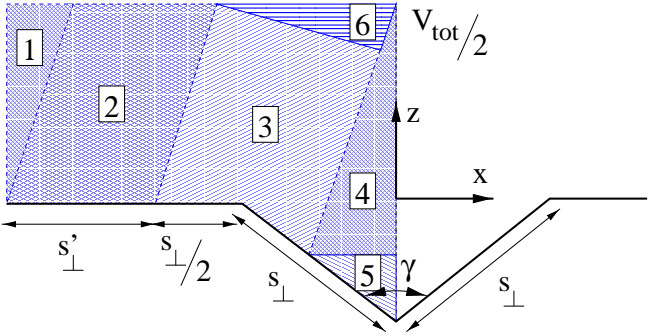}
\caption{A fluid is bounded by a substrate shaped as a wedge carved out of
an infinite planar wall.
The total volume of the fluid can be decomposed into the numbered
subvolumes that emerge naturally due to subvolumes $3$ and $5$ used for
describing a single ridge (first choice of the volume of integration
(\ref{subsub1})) and a single wedge (second choice of the volume of
integration
(\ref{subsub2})), respectively.}
\label{abra:ekfal}
\end{figure}
Due to the symmetry of the configuration we consider only half of the
wedge and focus on the case that $s_\perp$ is sufficiently large so that the
ridges and the wedge do not influence each other. One half
of the total volume of the fluid is decomposed into six numbered subvolumes
as shown in Fig. \ref{abra:ekfal}. We calculate the excess
adsorption [Eq. (\ref{gammahullam})] separately for each one of the
subvolumes by
adding and subtracting the profile of a fluid exposed to an infinite planar
wall $m_\pm^{\infty/2}$ in the integral of the order parameter as in
Subsecs. \ref{egywedgeridge} and \ref{sorozatapp}. 

For the first volume there is no line contribution resulting from the integral
over the difference of the actual profile $m_\pm$ from that in front of 
an infinite planar wall
$m_\pm^{\infty/2}$, because this difference is exponentially small in the
thermodynamic limit. There is also no surface term, because this first
volume touches the substrate only at one point.
However, following similar
considerations as for a single wedge or ridge, this volume
gives rise to a line contribution [see Eq. (\ref{altfelbontas})] 
to the excess adsorption [Eq. (\ref{gammahullam})]
due to the deviation of $m_\pm^{\infty/2}$ from $m_b$ with amplitudes:
\begin{equation}
\hat{\Gamma}^\pm_{l,1}(t,\gamma)=
\tan\frac{\pi-\gamma}{4}
\int_0^\infty  
\zeta[m^{\infty/2}_\pm(\zeta,t)-m_b(t)] d\zeta,
\label{elsoharomszogextra}
\end{equation}
where we have shifted the $z=const$ boundary to $+\infty$ with an exponentially
small correction to the integral.
 
In the thermodynamic limit ($s_\perp,\;s'_\perp\rightarrow\infty$) 
the adsorption profile $m_\pm$ in the second
 volume will tend exponentially to that of a fluid exposed to an infinite 
planar wall $m_\pm^{\infty/2}$. Thus it supplies a surface term with the
amplitudes
$\Gamma^\pm_s(t)=\Gamma^{\infty/2}_\pm (t)$ [see Eq. (\ref{adszdef})]
for the area $s'_\perp s_\parallel^{d-2}$ We
note that this subvolume two when extended to infinity in the $+z$ direction
overlaps with the other half of the wedge, but this results in only an
exponentially small correction to the excess adsorption, 
because the profile $m_\pm$ approaches the bulk value $m_b$ exponentially with
 the distance $\zeta$ from the wall. The subvolume two does not generate a
line contribution.

The third volume gives the same contributions [Eq. (\ref{altfelbontas})]
to the excess adsorption as a single
ridge of opening angle $(3\pi-\gamma)/2$ with the first choice of volume
of integration shown in Fig. \ref{abra:terf1}, i.e.,
a surface term with the amplitudes $\Gamma^\pm_s(t)=\Gamma^{\infty/2}_\pm (t)$ 
for an area $s_\perp s_\parallel^{d-2}$ [see Eq. (\ref{adszdef})] 
and line term amplitudes $\Gamma^\pm_{lr}(t,\gamma_r=(3\pi-\gamma)/2)$
[see Eq. (\ref{sorozatextra})]. Note that the volume
of integration can be extended to infinity in the direction parallel to the
bisector of the ridge even though in this case there is an overlap with the
other half of the wedge, because this results in only an
exponentially small correction to the excess adsorption 
as the profile $m_\pm$ approaches the bulk value
$m_b$ exponentially in this direction further away from the ridge.

Similarly to subvolume one, subvolume four does not generate a line term
resulting from the integral over the difference of the actual profile
$m_\pm$ from $m_\pm^{\infty/2}$. There is also no surface term, but similar
considerations as for a single wedge or ridge show that this subvolume gives
rise to a line contribution to the excess adsorption with amplitudes 
of the form:
\begin{eqnarray}
\hat{\Gamma}^\pm_{l,4}(t,\gamma)&=&
\left[\tan\frac{\pi-\gamma}{4}+\tan\frac{\gamma}{2}\right]
\nonumber\\&&\times
\int_0^\infty  
\zeta[m^{\infty/2}_\pm(\zeta,t)-m_b(t)] d\zeta,
\label{haromszogextra}
\end{eqnarray}
once we have shifted the $x=0$ boundary to $+\infty$ with an exponentially
small correction to the integral. 

The fifth subvolume together with its counterpart at $x>0$ gives 
the same contributions [Eq. (\ref{altfel2terf}] to the excess
adsorption as a wedge of opening angle $\gamma$ with the second choice of
the volume of integration shown in Fig. \ref{abra:terf2}(a), i.e., a surface
term with amplitudes
$\Gamma^\pm_s(t)=\Gamma^{\infty/2}_\pm (t)$ [see Eq. (\ref{adszdef})] for
the area $s_\perp s_\parallel^{d-2}$, and the 
line term amplitudes $\Gamma^\pm_{l,II}(t,\gamma)$ from Eq.
(\ref{altfel2terf}).

Finally the subvolume six yields only an exponentially small contribution to
the excess adsorption.

Adding up all contributions to the excess adsorption in this geometry
one obtains
\begin{widetext}
\begin{eqnarray}
&&\int_{V_{tot}}
d^dx\:m_\pm(\bm{x},t;\gamma)=V_{tot}m_b(t)+S_{tot}\Gamma_s^\pm (t)
+L\bigg[\Gamma_{lw}^\pm(t,\gamma_w)
+2\Gamma_{lr}^\pm(t,\gamma_r=(3\pi-\gamma_w)/2)
\nonumber\\
&&\left.+4\left(\frac{1}{\cos\frac{\gamma_w}{2}}-\frac{1}{\sin \gamma_w}
-\frac{1}{2}\tan\frac{\gamma_w}{2}\right)
\int_0^\infty  
\zeta[m^{\infty/2}_\pm(\zeta,t)-m_b(t)] d\zeta\right]
+{\cal{O}}(s_\perp^{-1}).
\end{eqnarray}
\end{widetext}
As in the case of a periodic array of wedges and ridges the line
contribution to the excess adsorption contains a combination of wedge and
ridge terms. Since different combinations thereof enter into the excess
adsorption of a periodic array and of a single embedded wedge, measurements
of both of them provide independent information. However, these
configurations do not allow one to access the ridge and wedge contributions
individually. In the case of a single embedded wedge the line contribution
to the excess adsorption contains in addition the first moment of the order
parameter profile of a semi-infinite planar system, which can be determined
independently from the knowledge of $m^{\infty/2}(\zeta)$.


\end{document}